\documentclass[paper]{JHEP3}

\usepackage{latexsym}
\usepackage{amsmath}
\usepackage{amscd}
\usepackage{amsfonts}
\usepackage{amsthm}
\allowdisplaybreaks[1]

\def\IN{\mathbb{N}}

\def\IR{\mathbb{R}}

\newcommand{\D}{\mathcal{D}}    
\newcommand{\Tr}{{\rm Tr}}      
\def\ap#1{\alpha^{\prime\,#1}}

\title{Higher order terms in the non-abelian D-brane effective action
and magnetic background fields}
\author{Alexander Sevrin and Alexander Wijns\thanks{Aspirant FWO}\\
    Theoretische Natuurkunde, Vrije Universiteit Brussel \\
    Pleinlaan 2, B-1050 Brussels, Belgium\\
        E-mail: \email{asevrin@tena4.vub.ac.be}, \email{awijns@tena4.vub.ac.be}}

\preprint{\hepth{0306260}}

\abstract{Recently a proposal for the non-abelian effective
D-brane action was given through order $\alpha'{}^4$. As the resulting
expressions turned out to be quite involved, some checks of this result
are called for. In the present paper we calculate the spectrum in
the presence of constant magnetic background fields and compare it to the
string theoretical result. Apart from a small typo in the original expression
(the overall sign of the $ \alpha '{}^4$ term),
we obtain perfect agreement. We discuss potential applications.}


\begin{document}
\section{Introduction}
The effective action for D-branes is one of the few tools
available for the study of the dynamics of D-branes. It is quite
surprising that, in the limit of slowly varying fields, the
effective action for a single D$p$-brane is known to all orders in
$ \alpha '$. It is given by the ten dimensional supersymmetric
Born-Infeld action dimensionally reduced to $p+1$ dimensions
\cite{BI}, \cite{susynbi}.

No such a result is presently available for the case of several,
say $n$, coinciding D$p$-branes. In leading order in $ \alpha '$,
it is the ten-dimensional $N=1$ supersymmetric $U(n)$ Yang-Mills
theory dimensionally reduced to $p+1$ dimensions \cite{witten}.
There are no ${\cal O}(\alpha')$ corrections. The bosonic ${\cal
O}(\alpha'{}^2)$ corrections were first obtained in \cite{direct}
and \cite{direct1}. The fermionic terms through this order were
obtained in \cite{goteborg} and \cite{bilal}. In \cite{goteborg},
supersymmetry fixed the correction while in \cite{bilal} a direct
calculation starting from four-point open superstring amplitudes
was used. Requiring the existence of certain BPS configurations,
called stable holomorphic bundles \cite{shm}, allows for a selfconsistent
determination of the effective action \cite{lies}. This was
applied in \cite{sk1} to determine the bosonic ${\cal
O}(\alpha'{}^3)$ terms in the effective action. In
\cite{groningen}, supersymmetry was used not only to confirm the
results of \cite{sk1} but to construct the terms quadratic in the
gauginos through this order as well. Later on, these results were
confirmed through a direct calculation of five point functions in
open superstring theory \cite{brasil}. Restricting to the special
case of four dimensions, one finds that, through this order, the
effective action also coincides with the one loop effective action
in $N=4$, $d=4$ super Yang-Mils \cite{sym1}, \cite{sym2},
\cite{sym3}.

Recently, the methods of \cite{lies} were used to determine the
effective action through order $ \alpha '{}^4$ \cite{us4}. Through
this order, the effective action is given by,
\begin{eqnarray}
{\cal L}=\frac{1}{g^2}\left({\cal L}_0+{\cal L}_2+{\cal L}_3+{\cal L}_4\right),
\end{eqnarray}
where the leading term is simply\footnote{Most of the time, we put
$2 \pi \alpha '=1$. Our metric follows the ``mostly plus'' convention.
The $u(n)$ generators are always
anti-hermitean and we use the following notation: $F^m\equiv
F_{a_1}{}^{a_2}F_{a_2}{}^{a_3}\cdots F_{a_m}{}^{a_1}
\equiv
F_{a_1a_2}F_{a_2a_3}\cdots F_{a_ma_1}\equiv F_{12}F_{23}\cdots
F_{m1}$.}
\begin{eqnarray}
\label{la0}
{\cal L}_0=\;=\; -
 \Tr\, \left\{\frac{1}{4}F^2\right\}\,.
\end{eqnarray}
Subsequently we have
\begin{eqnarray}
\label{la2}
{\cal L}_2=\mbox{STr} \left\{\frac{1}{8} F^4
- \frac{1}{32} F^2F^2
\right\},
\end{eqnarray}
where STr denotes the symmetrized trace prescription. At this
point both the overall multiplicative factor in front of the
action as well as the scale of the gauge fields got fixed
\cite{sk1}. The next term is\footnote{All results are of course
modulo field redefinition terms.}
\begin{eqnarray}
\label{la3}
{\cal L}_3=\frac{\zeta (3)}{2\pi^3}\Tr\left\{[D_3,D_2] D_4 F_{51} \, D_5 [D_4,D_3] F_{12} \right\}.
\end{eqnarray}
The overall coefficient of this term remained undetermined when
using the method of \cite{lies}. It was fixed by comparing it to
the partial result for this term in \cite{bilal1} which was
obtained by a direct string theoretic calculation. Note that this
expression is considerably simpler than the one which originally
appeared in \cite{sk1}. This is due to a different choice of basis
in which we express the action. Indeed, using partial integration,
Bianchi identities, ..., the action can be written in numerous
different ways. Finally the fourth order term is completely
determined by the method of \cite{lies} and it is given by
\cite{us4},
\begin{eqnarray}
{\cal L}_4 = {\cal L}_{4,0} + {\cal L}_{4,2} + {\cal L}_{4,4} \, ,\label{la4}
\end{eqnarray}
with
\begin{equation}
\begin{split}
{\cal L}_{4,0} & =  -\mbox{STr} \left( \frac{1}{12} F_{12}F_{23}F_{34}F_{45}F_{56}F_{61}
                 - \frac{1}{32} F_{12}F_{23}F_{34}F_{41}F_{56}F_{65}
                  + \frac{1}{384} F_{12}F_{21}F_{34}F_{43}F_{56}F_{65} \right) , \\
{\cal L}_{4,2} & = -\frac{1}{48} \mbox{STr} \Big( -2 \, F_{12}D_{1}D_{6}D_{5}F_{23}D_{6}F_{34}F_{45}
                                     - F_{12}D_{5}D_{6}F_{23}D_{6}D_{1}F_{34}F_{45}  \\
               & + 2 \, F_{12}\left[D_{6},D_{1}\right] D_{5}F_{23}F_{34}D_{4}F_{56}
                 + 3 \, D_{4}D_{5}F_{12}F_{23}\left[D_{6},D_{1}\right]F_{34}F_{56} \\
               & + 2 \, D_{6}\left[D_{4},D_{5}\right]F_{12}F_{23}D_{1}F_{34}F_{56}
                 + 2 \, D_{6}D_{5}F_{12}\left[D_{6},D_{1}\right]F_{23}F_{34}F_{45} \\
               & + 2 \, \left[D_{6},D_{1}\right]D_{3}D_{4}F_{12}F_{23}F_{45}F_{56} \\
               & + \left[D_{6},D_{4}\right]F_{12}F_{23}\left[D_{3},D_{1}\right]F_{45}F_{56} \Big) , \\
{\cal L}_{4,4} & = -\frac{1}{1440} \mbox{STr} \Big( D_6 [D_4,D_2]D_5 D_5 [D_1,D_3] D_6 F_{12} F_{34}
                   + 4 \, D_2 D_6 [D_4,D_1][D_5,[D_6,D_3]] D_5 F_{12} F_{34}  \\
                 & + 2 \, D_2 [D_6,D_4][D_6,D_1] D_5 [D_5,D_3] F_{12} F_{34}
                   + 6 \, D_2 [D_6,D_4]D_5[D_6,D_1][D_5,D_3] F_{12} F_{34} \\
                 & + 4 \, D_6 D_5 [D_6,D_4][D_5,D_1][D_4,D_3] F_{12} F_{23}
                   + 4 \, D_6 D_5 [D_4,D_2][D_6,D_1][D_5,D_3] F_{12} F_{34} \\
                 & + 4 \, D_6 [D_5,D_4][D_3,D_2][D_5,[D_6,D_1]] F_{12} F_{34} \\
                 & + 2 \, [D_6,D_1][D_2,D_6][D_5,D_4][D_5,D_3]F_{12}F_{34} \Big) \, .
\end{split}\label{la4e}
\end{equation}
The overall sign of $ {\cal L}_4$ is different from the one in
\cite{us4}. This is due to a typo in \cite{us4}.
Obviously, the expression for the ${\cal O}( \alpha '{}^4)$ terms
is very involved. So an independent check of these is called for.
In \cite{HT}, further developed in \cite{DST} and \cite{STT}, such
a test was proposed. One starts from two D$2p$-branes wrapped
around a $2p$-dimensional torus. When switching on constant
magnetic background fields, this yields, upon T-dualizing, two
intersecting D$p$-branes. String theory allows for the calculation
of the spectrum of strings stretching between different branes
\cite{callan}, \cite{leigh}. In the context of the effective
action, the spectrum should be reproduced by the mass spectrum of
the off-diagonal gauge field fluctuations. In \cite{test} it was
shown that the bosonic terms through ${\cal O}(\alpha'{}^3)$
correctly reproduce the spectrum of the gauge fields. In
\cite{test2}, the method was further extended such that the
fermionic terms could be tested as well. In the present paper we
turn to the test of the bosonic terms at order $ \alpha '{}^4$.
This is particularly interesting, as it is precisely at this order
that the mass spectrum such as obtained from the symmetrized trace
prescription for the non-abelian Born-Infeld \cite{ATSTR} (this
corresponds to $ {\cal L}_{4,0}$ in eq. (\ref{la4e})) starts to
deviate from the string theoretic spectrum \cite{HT}, \cite{DST},
\cite{STT}.

\section{The spectrum from string theory}
\label{string} We consider a constant magnetic background on two
coincident D$2p$-branes,
\begin{eqnarray}
{\cal F}_{2a-1\,2a}=i
\left(
   \begin{array}{cc}
     F_a & 0 \\
     0 & -F_a
   \end{array}
  \right),\label{mbg}
\end{eqnarray}
with $a\in\{1,2,\cdots,p\}$ and $F_a\in\IR$, $F_a>0$. We choose a
gauge such that $ {\cal A}_{2a-1}=0$, $\forall a$, and T-dualize
in the $2$, $4$, ..., $2p$ directions. We end up with two
intersecting D$p$-branes.  Taking the first brane located along
the 1, 3, ..., $2p-1$ directions, one finds that the other brane
has been rotated with respect to the first one over an angle
$\theta_1$ in the $12$ plane, over an angle $\theta_2$ in the $34$
plane, ..., over an angle $\theta_p$ in the $2p-1\,2p$ plane. The
angles are determined by the magnetic fields,
\begin{eqnarray}
\theta_a=2\,\arctan 2\pi \alpha ' F_a,\quad \forall a \in\{1,2,\cdots, p\}.\label{anglef}
\end{eqnarray}
One finds for the mass of the open strings stretching between
the two branes \cite{leigh}, \cite{HT}, \cite{DST},
\begin{eqnarray}
M^2= \frac{1}{2\pi \alpha '}\left(\sum_{b=1}^p(2\,m_b+1)\theta_b\pm 2\,\theta_a\right),\quad
a\in\{1,\cdots,p\},\quad m_b\in\IN.\label{spectrumstring}
\end{eqnarray}
In the previous, we temporarily reinstated the factors of $2\pi
\alpha '$.
\section{The spectrum from the effective action}

The mass formula given in eq. (\ref{spectrumstring}) should be
reproduced by the effective action. Taking the effective action
given in eqs. (\ref{la0}--\ref{la4e}), one turns on the magnetic
background given in in eq. (\ref{mbg}) and one subsequently
diagonalizes the linearized equations of motion for the
off-diagonal fluctuations. Expanding eq. (\ref{spectrumstring}) in
powers of $ \alpha '$ using eq. (\ref{anglef}) and setting $2\pi
\alpha '$ back to one, we get
\begin{eqnarray}
M^2=\sum_{b=1}^p2(2\,m_b+1)\left(F_b- \frac{F_b^3}{3}+ \frac{F_b^5}{5}\right)\pm 4
\left(F_a- \frac{F_a^3}{3}+ \frac{F_a^5}{5}\right)+ {\cal O}(F^7).\label{specexp}
\end{eqnarray}
{From} this it is clear that the terms linear in $F$ have to be
reproduced by $ {\cal L}_0$, those cubic in $F$ by $ {\cal L}_2$,
$ {\cal L}_3$ should not contribute to the spectrum and $ {\cal
L}_4$ is responsible for the terms quintic in $F$ in the spectrum.
\subsection{Leading order result}

We turn on a constant magnetic background $ {\cal F}_{ab}$ with
the corresponding background gauge potentials $ {\cal A}_a$. We
parameterize the gaugefields by $A_a= {\cal A}_a+ \delta A_a$. As
the calculation of the spectrum only probes $U(2)$ sub-sectors of
the full $U(n)$ theory \cite{spec2}, we take $U(2)$ as the gauge
group. We compactify $2p$ dimensions on a torus and introduce
complex coordinates for the compact directions,
$z^\alpha=(x^{2\alpha-1}\,-\,i\, x^{2\alpha})/\sqrt{2}$,
$\bar{z}^{\bar\alpha}=(z^\alpha)^*$, $\alpha\in\{1,\cdots,p\}$. We
take magnetic fields in the compact directions, such that $ {\cal
F}_{\alpha\beta}= {\cal F}_{\bar\alpha\bar\beta}=0$, ${\cal
F}_{\alpha\bar\beta}=0$ for $\alpha\neq\beta$ and\footnote{We do
not sum over repeated indices corresponding to {\em complex}
coordinates, unless indicated otherwise.}
\begin{eqnarray}
{\cal F}_{\alpha\bar\alpha}= i\left(
   \begin{array}{cc}
     f_\alpha & 0 \\
     0 & -f_\alpha
   \end{array}
  \right),
\end{eqnarray}
where the $f_\alpha$, $\alpha\in\{1,\cdots,p\}$ are imaginary
constants such that $if_\alpha=F_\alpha >0$. We are only
interested in the off-diagonal components of the gauge fields,
\begin{eqnarray}
\delta A=
i\left(
   \begin{array}{cc}
     0 & \delta  A^+ \\
     \delta  A^- & 0
   \end{array}
  \right),
\end{eqnarray}
as the diagonal fluctuations probe the abelian part of the action.
The spectrum for $ \delta A^+$ is equal to that of $ \delta  A^-$,
which reflects the two orientations of the strings stretching
between the two branes. Throughout the paper, we will investigate
the spectrum for $ \delta A^+$.

Linearizing the equations of motion which follow from $ {\cal
L}_0$ in eq. (\ref{la0}), we get,
\begin{eqnarray}
0&=&\left({\cal D}^2+4i
f_\alpha \right) \delta A^+_\alpha
-\sum_\beta {\cal D}_\alpha ({\cal
D}_\beta \delta A^+_{\bar{\beta}}+{\cal D}_{\bar{\beta}}\delta
A^+_\beta)\;, \nonumber\\
0&=&\left({\cal D}^2-4i
f_\alpha \right) \delta A^+_{\bar\alpha}
-\sum_\beta {\cal D}_{\bar\alpha }({\cal
D}_\beta \delta A^+_{\bar{\beta}}+{\cal D}_{\bar{\beta}}\delta
A^+_\beta)\;,
\label{Leading1}
\end{eqnarray}
where
\begin{eqnarray}
{\cal D}^2 \delta A^+= \left(\Box_{NC} + \sum_\beta{\cal D}_\beta {\cal
D}_{\bar{\beta}} +\sum_\beta{\cal D}_{\bar\beta} {\cal
D}_{{\beta}}\right),\label{Dkwad}
\end{eqnarray}
where $\Box_{NC}$ denotes the d'Alambertian in the non-compact
directions and we have
\begin{eqnarray}
{\cal D}_\alpha  \delta A^+=\left( \partial_\alpha +2i {\cal A}_\alpha \right) \delta  A^+,\quad
{\cal D}_{\bar\alpha}  \delta A^+=\left( \partial_{\bar\alpha} +2i {\cal A}_{\bar\alpha} \right) \delta  A^+.
\end{eqnarray}
Using
\begin{eqnarray}
{[} {\cal D}_ \alpha , {\cal D}_ {\bar\beta} {]}=2i \delta _{ \alpha \beta }f_ \alpha ,\label{Dcomm}
\end{eqnarray}
and choosing the gauge,
\begin{eqnarray}
\sum_ \beta ({\cal
D}_\beta \delta A^+_{\bar{\beta}}+{\cal D}_{\bar{\beta}}\delta
A^+_\beta)=0,
\end{eqnarray}
we can rewrite eq. (\ref{Leading1}) as
\begin{eqnarray}
0&=&\left(\Box_{NC}+2\sum_ \beta \left({\cal D}_ \beta {\cal D}_{\bar \beta }-i f_ \beta\right) +4i
f_\alpha \right) \delta A^+_\alpha , \nonumber\\
0&=&\left(\Box_{NC}+2\sum_ \beta \left({\cal D}_ \beta {\cal D}_{\bar \beta }-i f_ \beta\right) -4i
f_\alpha \right) \delta A^+_{\bar\alpha}.
\end{eqnarray}
In order to diagonalize this, we introduce a complete set of
functions on the torus,
\begin{eqnarray}
\phi _{\{m_1,m_2,\cdots,m_p\}}(z,\bar z)\equiv
{\cal D}^{m_1}_{z^1}{\cal D}^{m_2}_{z^2}\cdots{\cal D}^{m_p}_{z^p}
\phi _{\{0,0,\cdots ,0\}}(z,\bar z),
\end{eqnarray}
where $\phi _{\{0,0,\cdots ,0\}}$ is defined through,
\begin{eqnarray}
{\cal D}_{\bar \alpha }\,\phi _{\{0,0,\cdots ,0\}}(z,\bar z)=0,\qquad \forall\, \bar \alpha \in\{1,2,\cdots,p\}.
\label{grdst}
\end{eqnarray}
The function $\phi _{\{0,0,\cdots ,0\}}(z,\bar z)$ was explicitely
constructed in \cite{spec1} and \cite{spec2}. 
It is fully determined by eq. (\ref{grdst}) and the requirement that they satisfy proper boundary conditions.
Denoting the
non-compact coordinates collectively by $y$, we make the
expansion,
\begin{eqnarray}
\delta A^+(y,z,\bar z)=\sum_{(m_1,\cdots,m_p)\in\IN{}^{{}^p}} \delta A^{+\{m_1,\cdots ,m_p\}}(y)
\,\phi _{\{m_1,\cdots,m_p\}}(z,\bar z).
\end{eqnarray}
Using eq. (\ref{Dcomm}), one immediately gets,
\begin{eqnarray}
\left(\Box_{NC}-M^2\right) \delta A^{+\{m_1,\cdots,m_p\}}_{ \alpha }(y)=0,
\end{eqnarray}
with,
\begin{eqnarray}
M^2=2i \sum_{ \beta =1}^p(2m_ \beta +1)f_ \beta -4if_ \alpha ,
\end{eqnarray}
and
\begin{eqnarray}
\left(\Box_{NC}-M^2\right) \delta A^{+\{m_1,\cdots,m_p\}}_{\bar \alpha }(y)=0,
\end{eqnarray}
with,
\begin{eqnarray}
M^2=2i \sum_{ \beta =1}^p(2m_ \beta +1)f_ \beta +4if_ \alpha ,
\end{eqnarray}
which indeed agrees with the leading term in eq. (\ref{specexp}).
In the remainder of the paper, we will concentrate on the spectrum
of $ \delta A_{{ \alpha }}^+$ and denote it simply by $ \delta A_
\alpha $. It is a trivial exercise to extend the results to $
\delta  A_{{ \bar\alpha }}^+$.

\subsection{Lower order results}

For ${\cal L}_0 + {\cal L}_2$, the linearized
equations of motion become,
{\setlength\arraycolsep{2pt}
\begin{eqnarray}
0&=&\left[ (1+\frac{1}{3}f_\alpha^2-\frac{1}{6}\sum_\gamma
f_\beta^2){\cal D}^2+\frac{2}{3}\sum_\beta f_\beta^2 ({\cal
D}_\beta {\cal D}_{\bar{\beta}} -if_\beta -if_\alpha) +4i
(f_\alpha +\frac{2}{3} f_\alpha^3) \right]\delta A_\alpha \nonumber\\
&&-\sum_\beta \left[ 1+\frac{1}{3}(f_\alpha^2+ f_\beta^2)
-\frac{1}{6}\sum_\gamma f_\gamma^2 \right]{\cal D}_\alpha ({\cal
D}_\beta \delta A_{\bar{\beta}}+{\cal D}_{\bar{\beta}}\delta
A_\beta)\;, \label{L0+L1}
\end{eqnarray}
\noindent with ${\cal D}^2$ given in eq. (\ref{Dkwad}). As the
linearized equation of motion should be of the form
$(\Box_{NC}+\cdots) \delta A_ \alpha =0$, we need to make a field
redefinition,
\begin{eqnarray}
\delta \hat{A}_\alpha &=& \left(1+\frac{1}{3}f_\alpha^2
-\frac{1}{6}\sum_\beta f_\beta^2\right)\delta A_\alpha\;.\label{fr2}
\end{eqnarray}
Using this and eq. (\ref{Dcomm}), we can rewrite eq. (\ref{L0+L1}) as
\begin{eqnarray}
0&=&\left[ \Box_{NC} + 2 \sum_\beta (1+ \frac{1}{3}f_\beta^2)({\cal
D}_\beta {\cal D}_{\bar{\beta}} -if_\beta)+ 4i(f_\alpha +
\frac{1}{3}f_\alpha^3) \right] \delta \hat{A}_\alpha \nonumber\\
&&-\frac 2 9 \left[ (f_ \alpha ^2-\frac 1 2 \sum_ \gamma f_ \gamma
^2)\sum_ \beta f_ \beta ^2( {\cal D}_ \beta {\cal D}_{\bar \beta
}-if_ \beta )+2if_ \alpha ^3(f_ \alpha ^2-\frac 1 2 \sum_ \beta f_
\beta ^2)
\right]\delta A_ \alpha \nonumber\\
&&-\sum_\beta \left[ 1+\frac{1}{3}(f_\alpha^2+ f_\beta^2)
-\frac{1}{6}\sum_\gamma f_\gamma^2 \right]{\cal D}_\alpha ({\cal
D}_\beta \delta A_{\bar{\beta}}+{\cal D}_{\bar{\beta}}\delta
A_\beta)\; .
\label{eom}
\end{eqnarray}
\noindent The first line is precisely what we need. Indeed,
proceeding as in the previous section, one finds that the spectrum
of $ \delta \hat A_ \alpha $ reproduces eq. (\ref{specexp})
through order $F^3=(i\,f)^3$. The second line in eq. (\ref{eom})
can presently be ignored as it will contribute order $f^5$
corrections to the spectrum. However, these terms will interfere
with the contributions arising from $ {\cal L}_4$ (see the
analysis in the next section). Finally, the last line of eq.
(\ref{eom}) can be eliminated by making an appropriate gauge
choice,
\begin{eqnarray}
\sum_\beta (1+\frac{1}{3}f_\beta^2)\left({\cal D}_\beta \delta
A_{\bar{\beta}}+{\cal D}_{\bar{\beta}}\delta A_\beta\right) &=&
0\;.\label{gauge}
\end{eqnarray}
Note that this again yields terms which should be taken into
account when analyzing the $ {\cal L}_4$ contributions.

The $\alpha'^3$ term, eq. (\ref{la3}), results in the following
linearized equations of motion,
\begin{eqnarray}
0&=&-\frac{4 \zeta(3)}{\pi^3}\left[ \left(f_\alpha^2 {\cal D}^2
-2\sum_\beta f_\beta^2({\cal D}_\beta {\cal D}_{\bar{\beta}}
-if_\beta)\right)\left( {\cal D}^2+4if_\alpha \right)
\right]\delta A_\alpha \nonumber\\
&& + \frac{4 \zeta(3)}{\pi^3}\sum_{\beta}{\cal
D}_\alpha\left[ (f_\alpha^2 + f_\beta^2){\cal D}^2- 4\sum_\gamma f_\gamma^2
({\cal D}_\gamma {\cal D}_{\bar{\gamma}} -if_\gamma) \right]({\cal
D}_\beta \delta A_{\bar{\beta}}+{\cal D}_{\bar{\beta}}\delta
A_\beta)\;. \label{L3}
\end{eqnarray}
\noindent Using eq. (\ref{L3}) one can see that the first line of
eq. (\ref{eom}) still holds for ${\cal L}_0 + {\cal L}_2 + {\cal
L}_3$, if we now take $\delta \hat{A}_\alpha$ to be,
\begin{eqnarray}
\delta \hat{A}_\alpha &=& \left[1+\frac{1}{3}f_\alpha^2\left( 1-
\frac{12\zeta (3)}{\pi^3}{\cal D}^2 \right) -\frac{1}{6}\sum_\beta
f_\beta^2 \left( 1- \frac{48\zeta (3)}{\pi^3}({\cal D}_\beta {\cal
D}_{\bar{\beta}} -if_\beta) \right) \right]\delta
A_\alpha\;,\nonumber\\
\end{eqnarray}
\noindent while modifying the gauge condition (\ref{gauge}) to,
\begin{eqnarray}
\sum_\beta \left[1+\frac{1}{3}f_\beta^2\left(1 - \frac{12\zeta
(3)}{\pi^3}{\cal D}^2\right) \right]\left({\cal D}_\beta \delta
A_{\bar{\beta}}+{\cal D}_{\bar{\beta}}\delta A_\beta\right) &=&
0\;.\label{gauge2}
\end{eqnarray}
\noindent This will introduce additional terms in the spectrum of
order $f^6$ which will interfere with contributions coming from $
{\cal L}_5$. As the analysis of the present paper is limited to $
{\cal L}_4$ ($ {\cal L}_5$ is not even known), we can safely
ignore them.

Concluding, we find that ${\cal L}_0+ {\cal L}_2 +{\cal L}_3$
correctly reproduces the spectrum, eq. (\ref{specexp}), through
this order.

\subsection{The order $\alpha '{}^4$ result}
We now turn to the main point of the present paper: the
contributions to the spectrum which arise from $ {\cal L}_4$.

As the order increases, the calculations become rather
tedious, one of the reasons being the symmetrized trace
prescription. It turns out, however, that in our particular case
one can very easily reduce the symmetrized trace to an ordinary
trace.

Since we are only interested in the linearized form of the
equations of motion, we only need to consider the symmetrized
trace of a product of matrices, of which at most two have off-diagonal 
components\footnote{In case only one of them is off-diagonal, 
the operations STr and Tr coincide, so we only consider
the other case.}. A convenient way to do this was proposed in
\cite{DST}. For our purpose, their more general formula simplifies
in the following way: consider a product of $2n$ abelian
fieldstrengths $ {\cal F}_m$, $m\in\{1,\cdots,2n\}$ given by,
\begin{eqnarray}
{\cal F}_{m}= i\left(
   \begin{array}{cc}
     F_m & 0 \\
     0 & -F_m
   \end{array}
  \right),
\end{eqnarray}
and two arbitrary two by
two matrices with only off-diagonal components, which we call $G$
and $H$, 
\begin{eqnarray}
G= i\left(
   \begin{array}{cc}
     0 & G^+ \\
     G^- & 0
   \end{array}
  \right),
\qquad
H= i\left(
   \begin{array}{cc}
     0 & H^+ \\
     H^- & 0
   \end{array}
  \right).
\end{eqnarray}
Then we have,
\begin{eqnarray}
\mbox{STr}\left(GH{\cal F}_1 \cdots {\cal F}_{2n} \right) &=&
\frac{(-1)^n}{2n+1} F_1\cdots F_{2n}\Tr (GH)\;.
\end{eqnarray}
We see that in this case, taking a symmetrized trace is no more
difficult than taking an ordinary trace.

Using this result, the linearized equations of motion coming from
${\cal L}_{4,0}$ are still easy to obtain and are given by,
\begin{eqnarray}
\label{purestr}
0 &=& -\frac 1 5 \left[ \sum_\beta \left( \frac 1 4 f_\beta^4 + \frac 1 2 f_\alpha^2
f_\beta^2 - \frac 1 8 f_\beta^2 \sum_\gamma f_\gamma^2 \right)\D^2
 + if_\alpha\sum_\beta \left( 4f_\alpha^2 f_\beta^2
+ f_\beta^4 - \frac 1 2 f_\beta^2 \sum_\gamma f_\gamma^2 \right) \right.\nonumber\\
&& -\left. 2\sum_\beta \left( f_\beta^4 + f_\alpha^2 f_\beta^2 - \frac 1 2 f_\beta^2
\sum_\gamma f_\gamma^2\right) (\D_\beta
\D_{\bar\beta} -if_\beta) - f_\alpha^4\D^2 - 12if_\alpha^5
\right]\delta A_\alpha \nonumber \\
&& - \frac 1 5 \sum_\beta \left[ f_\alpha^4 +
f_\beta^4 + f_\alpha^2 f_\beta^2 - \frac 1 2 \sum_\gamma \left( \frac 1
2 f_\gamma^4 +  f_\alpha^2 f_\gamma^2 + f_\beta^2 f_\gamma^2 - \frac 1
4 f_\gamma^2 \sum_\delta f_\delta^2\right) \right]\times \nonumber \\
&& \D_\alpha \left({\cal D}_\beta \delta A_{\bar{\beta}}+{\cal D}_{\bar{\beta}}\delta
A_\beta\right).
\end{eqnarray}
\noindent This expression would lead to a correction to the mass
spectrum which is easily shown to deviate from eq. (\ref{specexp}).
This explicitely demonstrates, as was known long before, \cite{HT}, \cite{DST}, that from this order on,
the symmetrized trace prescription should receive
corrections. After quite a lengthy calculation we
find the linearized equations of motion for ${\cal L}_{4,2}+{\cal
L}_{4,4}$ to be,
\begin{eqnarray}
\label{derterms}
0 &=& \left\{ \Bigg[
\frac{1}{180} if_\alpha^3 \D^2 + \frac{4}{15}f_\alpha^4 + \sum_\beta \Bigg( \frac{1}{45}if_\alpha f_\beta^2
\D^2 - \frac{7}{45} if_\alpha f_\beta^2 (\D_\beta \D_{\bar\beta} -if_\beta) - \frac{7}{90}f_\beta^4 -
\frac 1 5 f_\alpha^2 f_\beta^2\right.\nonumber\\
&& \left. \left.+\frac{1}{18} f_\beta^2 \sum_\gamma f_\gamma^2\right) \right]\D^2 +
\sum_\beta \left( \frac 4 9 f_\alpha^2 f_\beta^2 + \frac{4}{45} f_\beta^2
\sum_\gamma f_\gamma^2 \right)(\D_\beta \D_{\bar\beta} -if_\beta) \nonumber \\
&& \left.+ \frac 4 5 if_\alpha^5 - if_\alpha \sum_\beta \left(\frac{14}{45}f_\beta^4 -
\frac{8}{15}f_\alpha^2 f_\beta^2  - \frac 2 9 f_\beta^2 \sum_\gamma f_\gamma^2 \right)
\right\}\delta A_\alpha \nonumber\\
&&- \D_\alpha \sum_\beta \left[ \left( \frac{1}{180} i f_\alpha^3 - \frac{1}{18} if_\alpha
f_\beta^2 + \frac{1}{36}if_\alpha \sum_\gamma f_\gamma^2
\right)\D^2 -\frac{2}{45}i f_\alpha \sum_\gamma f_\gamma^2(\D_\gamma \D_{\bar\gamma}
-if_\gamma)\right.\nonumber \\
&& \left. -\frac{4}{45} f_\alpha^2 f_\beta^2 -\frac{4}{45} \sum_\gamma \left(
f_\gamma^4 - f_\beta^2 f_\gamma^2 + f_\alpha^2 f_\gamma^2 \right)
\right]\left({\cal D}_\beta \delta A_{\bar{\beta}}+{\cal
D}_{\bar{\beta}}\delta A_\beta\right)\nonumber\\
&& + \D_\alpha \sum_\beta \left( \frac{1}{360} f_\alpha
f_\beta \D^2 - \frac{1}{18}if_\alpha^2 f_\beta +
\frac{1}{180}if_\beta^3 + \frac{1}{36}if_\beta
\sum_\gamma f_\gamma^2\right)\D^2 \left({\cal D}_\beta \delta
A_{\bar{\beta}}-{\cal D}_{\bar{\beta}}\delta
A_\beta\right).\nonumber \\
\end{eqnarray}

\noindent The sum of the righthand sides of eqs. (\ref{purestr})
and (\ref{derterms}), together with the contributions coming from
the second line of eq. (\ref{eom}) and those arising from the gauge choice,
eq. (\ref{gauge2}),
should now produce the correct
$F^5$ terms in eq. (\ref{specexp}). Careful analysis shows that
the linearized equations of motion corresponding to the total
lagrangian, eqs. (\ref{la0}--\ref{la4e}), can eventually be
written as,

\begin{eqnarray}
\left[ \Box_{NC} + 2 \sum_\beta (1+ \frac{1}{3}f_\beta^2 +
\frac{1}{5}f_\beta^4)({\cal D}_\beta {\cal D}_{\bar{\beta}}
-if_\beta)+ 4i(f_\alpha + \frac{1}{3}f_\alpha^3 +
\frac{1}{5}f_\alpha^5) \right] \delta \hat{A}_\alpha &=& 0\,,\quad
\label{eom4}
\end{eqnarray}

\noindent where, again, terms at fifth and higher order are
ignored. The total correction to the eigenvectors $\delta
A_\alpha$, which appear in eq. (\ref{eom4}), is given by,

\begin{eqnarray}
\delta \hat{A}_\alpha &=& \left[1+\frac{1}{3}f_\alpha^2\left(
1-\left( \frac{12\zeta (3)}{\pi^3} - \frac{1}{60}if_\alpha\right)
{\cal D}^2 + \frac{22}{15}f_\alpha^2 - \frac{1}{30}\sum_\gamma
f_\gamma^2
\right)\right.\nonumber \\
&&\left. -\frac{1}{6}\sum_\beta f_\beta^2 \Bigg( 1-
\left(\frac{48\zeta (3)}{\pi^3}-\frac{14}{15}if_\alpha
\right)\left({\cal D}_\beta {\cal D}_{\bar{\beta}} -if_\beta
\right) - \frac{2}{15}if_\alpha {\cal D}^2
\right.\nonumber\\
&&\left. \left. + \frac{23}{30}f_\beta^2 -
\frac{29}{60}\sum_\gamma f_\gamma^2\right) \right]\delta A_\alpha
+ \sum_\beta {\cal D}_\alpha \Bigg( \frac{1}{360} f_\alpha f_\beta
{\cal D}^2 - \frac{1}{18}if_\alpha^2 f_\beta \nonumber\\
&&+ \frac{1}{180}if_\beta^3 + \frac{1}{36}if_\beta \sum_\gamma
f_\gamma^2 \Bigg)\left( {\cal D}_\beta \,\delta A_{\bar{\beta}} -
{\cal D}_{\bar{\beta}} \,\delta A_\beta \right)\;.\label{fields4}
\end{eqnarray}

\noindent The gauge condition (\ref{gauge2}) also gets fourth
order contributions and becomes,

\begin{eqnarray}
\sum_\beta \left[1+\frac{1}{3}f_\beta^2\left(1 - \frac{12\zeta
(3)}{\pi^3}{\cal D}^2 + \frac{2}{15}\sum_\gamma f_\gamma^2\right)
+ \frac{1}{5} f_\beta^4 \right]\left({\cal D}_\beta \delta
A_{\bar{\beta}}+{\cal D}_{\bar{\beta}}\delta A_\beta\right) &=&
0\;.\quad \label{gauge3}
\end{eqnarray}

Eq. (\ref{eom4}) exactly leads to the mass spectrum in eq.
(\ref{specexp}). This shows that, if we redefine the mass
eigenvectors $\delta A_\alpha$ in an appropriate way and impose
the right gauge condition, we obtain total agreement with string
theoretical calculations up to fourth order in $\alpha '$!

\section{Discussion}
The non-abelian D-brane effective action is known through order $ \alpha '{}^4$, \cite{us4}.
In the present paper, we performed a successful test of this result. Indeed when switching on 
constant magnetic background fields, we showed that through this order, the spectrum agrees with the one 
obtained from a direct string theoretical calculation. The contributions coming from the symmetrized trace part of
the lagrangian, ${ {\cal L}}_{4,0}$, combined with those arising from the derivative terms in the action,
$ {\cal L}_{4,2}+ {\cal L}_{4,4}$, and those which arose from $ {\cal L}_2$ as a consequence of the 
field redefinition and the gauge choice, precisely reproduce the $ \alpha '^4$ terms in the spectrum, eq. (\ref{specexp}).
However, we would like to stress that this does not check every coefficient in the action. Indeed, when going through
the details of the calculation, one finds {\em e.g.} that 
the last term in $ {\cal L}_{4,2}$ and the second and the last term in $ {\cal L}_{4,4}$ 
do not contribute at all. 

Nonetheless, combining this
test with the fact that the calculation of $ \alpha '{}^4$ term in \cite{us4} required solving $1816$ algebraic 
equations in $546$ unknowns yielding a unique solution, shows that we can be very confident about the results in \cite{us4}.

Yet another test is provided by the results in \cite{mees3} (see also \cite{wyllard}). Requiring supersymmetry, 
the derivative terms in the abelian theory were determined through order $ \alpha '{}^4$. While this method
does not fix the overall constant in front of these terms (they form an independent supersymmetry invariant), the relative
coefficients are fixed. Taking the abelian limit of eq. (\ref{la4e}) gives a result which perfectly agrees with the one
in \cite{mees3}, however in our case the overall coefficient is fixed. 

Eqs. (\ref{la3}) and (\ref{la4e}) are very involved. At first sight there seems to be little hope that a closed expression to all
orders in $ \alpha '$ can be found. The possibility of making field redefinitions further complicates matters. We are convinced
that as a first step, the derivative corrections in the abelian limit should be investigated. If there is any organizational 
principle for the non-abelian effective action to all orders in $ \alpha '$, this should be true for the full abelian effective 
action, which obviously is much simpler, as well. This is presently being studied.

Finally, in \cite{tokyo}, the recombination of intersecting D$1$-branes was analyzed using the leading term in the 
non-abelian D-brane effective action by studying the tachyonic configurations. 
While the analysis of \cite{tokyo} is performed in a gauge different from ours, it is straightforward using eqs. (\ref{eom})
and (\ref{fr2}) to repeat their analysis through second order. No essential new features are added to their conclusions. However,
from third order on, the field redefinitions are more subtle as they involve derivative terms as well.
So it would be interesting, after T-dualizing the results given in previous sections, to study higher order effects
on D-string recombination along the lines of \cite{tokyo} including the corrections through order
$ \alpha '{}^4$. 
\bigskip

\acknowledgments

\bigskip

The authors are supported in part by the ``FWO-Vlaanderen''
through project G.0034.02, in part by the Federal Office for
Scientific, Technical and Cultural Affairs through the
Interuniversity Attraction Pole P5/27 and in part by the European
Commission RTN programme HPRN-CT-2000-00131, in which they are
associated to the University of Leuven. We thank 
Korneel van den Broek and in particular Paul Koerber for 
useful discussions. When finishing this paper, we became aware of a 
related check, the study of the spectrum of intersecting D1-branes (along the lines
suggested in \cite{tokyo}), being performed by Satoshi Nagaoka \cite{sat}. We thank the author for
friendly and interesting correspondence.

\end{document}